\documentclass[graybox]{svmult}

\usepackage{mathptmx}       % selects Times Roman as basic font
\usepackage{helvet}         % selects Helvetica as sans-serif font
\usepackage{courier}        % selects Courier as typewriter font
\usepackage{type1cm}        % activate if the above 3 fonts are
\usepackage{makeidx}         % allows index generation
\usepackage{graphicx}        % standard LaTeX graphics tool
\usepackage{multicol}        % used for the two-column index
\usepackage[bottom]{footmisc}% places footnotes at page bottom

\usepackage{amsmath,amsfonts,amssymb,mathrsfs}

\makeindex             % used for the subject index
\begin{document}

\title*{Gravity and Scalar Fields}
\author{Thomas P. Sotiriou}
\institute{School of Mathematical Sciences \& School of Physics and Astronomy, University of Nottingham, University Park, Nottingham, NG7 2RD, UK}
\maketitle

\abstract*{Gravity theories with non-minimally coupled scalar fields are used as characteristic examples in order to demonstrate the challenges, pitfalls and future perspectives of considering alternatives to general relativity. These lecture notes can be seen as an illustration of concepts, subtleties and techniques present in all alternative theories, but they also provide a brief review of generalised scalar-tensor theories. }

\abstract{Gravity theories with non-minimally coupled scalar fields are used as characteristic examples in order to demonstrate the challenges, pitfalls and future perspectives of considering alternatives to general relativity. These lecture notes can be seen as an illustration of features, concepts and subtleties that are present in most types of alternative theories, but they also provide a brief review of generalised scalar-tensor theories. }

\section{Introduction}
\label{sec:1}

The predictions of general relativity are in impressive agreement with experiments whose characteristic length scale ranges from microns ($\mu m$) to about an astronomical unit (AU). On the other hand, the theory is expected to break down near the Planck length, $l_p\approx 1.6\times 10^{-35}m$, and a quantum theory of gravity is needed in order  to adequately describe phenomena for which such small length scales are relevant. There are really no gravitational experiments that give us access to the region between the Plack length and the micron, so one has to admit that we have no direct evidence about how gravity behaves in that region.\footnote{However, one can infer certain properties of gravity indirectly. Matter couples to gravity and we understand and probe the structure and behaviour of particles and fields at scales much smaller than the micron, so if one is given a model that describes how gravity interacts with matter then one could in principle gain insight into some aspects of gravity through the behaviour of matter. Applying this logic to the quantum aspects of gravity has given rise to what is called Quantum Gravity Phenomenology \cite{AmelinoCamelia:2004hm,Liberati:2013xla}. The fact that the gravitational coupling is very weak poses a particular challenge in such an approach, but smoking gun signals can still exist in certain models.} 

It was perhaps much more unexpected that experiments probing length scales much larger than the solar system held surprises related to gravity. General relativity can only fit combined cosmological and galactic and extragalactic data well if there is a non vanishing cosmological constant and about 6 times more Dark Matter --- matter which we have so far detected only through its gravitational interaction --- than visible matter (see, for instance, Ref.~\cite{Ade:2013zuv}). Moreover, the value of the cosmological constant has to be very small, in striking disagreement with any calculation of the vacuum energy of quantum fields, and mysteriously the associated energy density is of the same order of magnitude as that of matter currently \cite{Weinberg:1988cp,Carroll:2000fy}. These puzzles have triggered the study of dynamical Dark Energy models, that come to replace the cosmological constant.

Since general relativity is not a renormalizable theory, it is expected that deviation from it will show up at some scale between the Planck scale and the lowest length scale we have currently accessed. It is tempting to consider a scenario where those deviation persist all the way to cosmological scales and account for Dark Matter and/or Dark Energy. After all, we do only detect these dark component through gravity. However, there is a major problem with this way of thinking. There is no sign of these modifications in the range of scales for which we have exhaustively tested gravity. So, they would have to be relevant at very small scales, then somehow switch off at intermediate scales, then switch on again at larger scales.  It is hard to imagine what can lead to such behaviour, which actually contradicts our basic theoretical intuition about separation of scales and effective field theory. Nonetheless, intuition is probably not a good enough reason to not rigorously explore an idea that could solve two of the major problem of contemporary physics at once. This explain the considerable surge of interest in alternative theories of gravity in the last decade or so.

Considering alternatives to a theory as successful as general relativity can be seen as a very radical move. However, from a different perspective it can actually be though of as a very modest approach to the challenges gravity is facing today. Developing a fundamental theory of quantum gravity from first principle and reaching the stage where this theory can make testable predictions has proved to be a very lengthy process. At the same time, it is hard to imagine that we will gain access to experimental data at scales directly relevant to quantum gravity any time soon. Alternative theories of gravity, thought of as effective field theories, are the phenomenological tools that provide the much needs contact between quantum gravity candidates and observations at intermediate and large scales. 

The scope of these notes is to briefly review the challenges one in bound to face when considering alternatives to general relativity and discuss various ways to overcome (some of) them. Instead of providing rigorous and general but lengthy arguments, I will mostly resort to the power of examples. The examples will be based on gravity theories with additional scalar degrees of freedom, so these notes will also act as a brief review of generalised scalar-tensor theories and their properties.

I have made extensive reference to various length scales in the arguments presented so far and one can rightfully feel uncomfortable talking about length scales when it comes to gravity. The strength of the gravitational interaction has to do with curvature and lengths are not even invariant under coordinate transformations. Indeed, the Planck length can only be understood as a fundamental invariant length as the inverse of the square root of a fundamental curvature scale (which has dimensions of 1 over a length square). In this spirit, it would be preferable to talk about the range of curvatures in which we have tested gravity. Actually, the experiments that span the range of lengths $\mu m$ - AU all lie in a very narrow band of curvatures. This is not so surprising, as they are all weak-field experiments. This applies to binary pulsars as well as, even though the two companions that form the binary are compact enough to exhibit large curvatures in their vicinity, the gravitational interaction between them is still rather weak as the two stars are not close enough to be in the region of strong curvature. Hence, if we think in terms of curvatures, the range in which we have tested general relativity appears even more restricted. Neutron stars and stellar and intermediate mass black holes can exhibits curvatures which are many orders of magnitudes larger than the usual weak-field experiments. It is, therefore, particularly interesting to understand the structure of such objects and the phenomena that take place in their vicinity in alternative theories of gravity. They are most likely the new frontier in gravitational physics.

The rest of these notes is organised as follows: In section \ref{sec:2} I  lay out the basic assumption of general relativity and very briefly (and intuitively) discuss the consequences of relaxing these assumptions. The main scope of this section is to give an idea of what alternative theories of gravity are about and what kind of  problems one usually faces when deviating from general relativity. In sections \ref{sec:3} and \ref{sec:4} I  attempt to support the statements made in the previous section by considering characteristics examples from (generalised) scalar-tensor gravity theories. Section \ref{sec:5} focuses on black hole physics in scalar-tensor gravity. Section \ref{conc} contains  conclusions.

\section{General Relativity and beyond}
\label{sec:2}

\subsection{General Relativity: Basic assumptions and uniqueness}

The action of general relativity is
\begin{equation}
\label{EHaction}
S=\frac{1}{16 \pi G}\int d^4 x \sqrt{-g} (R-2 \,\Lambda) + S_m (g_{\mu\nu},\psi)\,,
\end{equation}
where $G$ is Newton's constant, $g$ is the determinant of the spacetime metric $g_{\mu\nu}$, $R$ is the Ricci scalar of the metric, $\Lambda$ is the cosmological constant, and $S_m$ is the matter action. $\psi$ collectively denotes the matter fields, which are understood to couple minimally to the metric. 

Coupling the matter fields $\psi$ only to the metric and with the standard prescription of minimal coupling guaranties that the Einstein Equivalence Principle is satisfied. That is, test particles follow geodesics of the metric and non-gravitational physics is locally Lorentz invariant and position invariant \cite{willbook}. The reason why the  last two requirements are satisfied once matter is minimally coupled is that in the local frame the metric is flat to second order in a suitably large neighbourhood of a space-time point and $S_m$ reduces to the action of the Standard Model. It is worth elaborating a bit more on universality of free fall and how this is related to the form of the matter action. 

Consider the stress-energy tensor $T_{\mu\nu}$ of a pressure-less fluid, usually referred to as dust. An infinitesimal volume element of such a fluid is as close as one can get to a test particle. A rather straightforward calculation reveals that the conservation of the stress-energy tensor, $\nabla^\mu T_{\mu\nu}=0$, implies that the 4-velocity of the fluid satisfies the geodesic equation. That is, $\nabla^\mu T_{\mu\nu}=0$ implies that test particles follow geodesics. On the other hand, the conservation of the stress-energy tensor can be shown to follow from diffeomorphism invariance of the matter action $S_m$, {\em provided} that the matter fields are on shell (they satisfy their field equations). 

Let $\xi^\mu$ be the generator of a diffeomorphism and ${\cal L}_\xi$ denote the associated Lie derivative. Diffeomorphism invariance of the matter action implies
\begin{equation}
{\cal L}_\xi S_m=0\,.
\end{equation}
One can express the action of the Lie derivative in terms of functional derivatives of $S_m$ with respect to the fields, {\em i.e.}
\begin{equation}
\frac{\delta S_m}{\delta g^{\mu\nu}} {\cal L}_\xi g^{\mu\nu} +\frac{\delta S_m}{\delta \psi}{\cal L}_\xi \psi=0\,.
\end{equation}
However, $\delta S_m/\delta \psi=0$ are actually the field equations for $\psi$. So, on shell we have
\begin{equation}
\frac{\delta S_m}{\delta g^{\mu\nu}} {\cal L}_\xi g^{\mu\nu}=0\,.
\end{equation}
With the usual definitions for the stress-energy tensor and for the action of a Lie derivative on the metric and after some manipulations, the above equation can take the form
\begin{equation}
\int d^4 x \sqrt{-g} T_{\mu\nu} \nabla^\mu \xi^\nu=0\,.
\end{equation}
Finally, integrating by parts and taking into account that $\xi^\mu$ vanishes at the boundary yields
\begin{equation}
\label{diffeolast}
\int d^4 x \sqrt{-g} (\nabla^\mu T_{\mu\nu})\xi^\nu=0\,.
\end{equation}
Since, $\xi^\mu$ is a generic diffeomorphism, eq.~(\ref{diffeolast}) implies that $\nabla^\mu T_{\mu\nu}=0$.

In conclusion, diffeomorphism invariance of the matter action allows one to link geodesic motion with the requirement that the matter fields are on shell. An important assumption here is that there is no field other than the metric that couples to the matter fields $\psi$ and at the same time enters the gravitational action as well. This assumption is reflected in the condition that $\delta S_m/\delta \psi=0$, {\em i.e.}~{\em all} fields other than the metric are on shell.  If there were a field, say $\phi$ entering both $S_m$ and the gravitational action, then $\delta S_m/\delta \phi=0$ would not actually be its field equation and it would not be sensible to impose it as a condition by assuming that this field is on shell.

One more point that is worth stressing is that in the arguments and calculations shown above one only makes reference to the matter action. This implies that they are not specific to general relativity. Instead, they will apply to any theory in which the matter couples only to the metric through minimal coupling.

In conclusion, the requirement to satisfy the Einstein Equivalence Principle, which has been experimentally tested to very high accuracy, pins down the matter action and the coupling between matter and gravity.
What is left is to argue why the dynamics of $g_{\mu\nu}$ should be governed by the first integral in eq.~(\ref{EHaction}), known as the Einstein--Hilbert action. Luckily, this requires less work as Lovelock has provided us with a theorem \cite{Lovelock:1971yv,Lovelock:1972vz} stating that this is indeed the {\em unique} choice, provided that the following assumptions hold true:
\begin{enumerate}
\item The action is diffeomorphism-invariant;
\item it leads to second-order field equations for the metric;
\item we are restricting our attention to 4 dimensions;
\item no fields other than the metric enter the gravitational action.
\end{enumerate}

\subsection{Less assumptions means more degrees of freedom!}
\label{lessassum}

We now consider what would be the implications of giving up one of the assumptions listed above. Let us start by relaxing the assumption that the gravitational action depends only on the metric, and allow a dependence on a new field $\phi$. Obviously, we would need to dictate how the gravitational action depends on $\phi$ in order to pin down the theory we are considering. However, as it should be clear from the analysis in the previous section, if we were to allow this new field to enter the matter action and couple to the matter fields then we would have violations of the Einstein Equivalence Principle and signatures of this coupling would appear in non-gravitational experiments. Constraints on universality of free fall, local Lorentz symmetry in the matter sector, and deviations from the standard model in general are orders of magnitude more stringent than constraints coming from gravitational experiments. This explains why in the literature the common approach is to assume that any new fields do not enter the matter action, or at least that the coupling between these field and matter is weak enough to be irrelevant at low energies. We will follow the same line of thought in what comes next. It should, however, be clear that if there are new fields in the gravity sector at the classical level, then one would expect that quantum corrections will force them to couple to the matter fields. Hence, a consistent theory should actually include a mechanism that naturally suppressed the coupling between these new fields and matter. This is required in order to theoretical justify what phenomenologically seems to be the only option.

A thorny issue is that of field redefinitions. Note that all of the assumptions, conditions, and requirements discussed above, should in principle be posed as ``there exists a choice of fields where...''. This becomes particularly relevant when one has extra fields mediating gravity. Suppose, for instance, that $\phi$ did couple to matter but in such a way that I can introduce a new metric, $\tilde{g}_{\mu\nu}$, which can be given in closed form in terms of $g_{\mu\nu}$ and $\phi$ (and potentially its derivatives), so that matter actually couples minimally to $\tilde{g}_{\mu\nu}$. Then, the whole theory can be re-written in terms of $\tilde{g}_{\mu\nu}$ and $\phi$ and the matter action will be the conventional one with matter coupling only to a metric with minimal coupling. 

What would now happen if we kept the field content unchanged and we instead relaxed any of the other 3 assumptions of Lovelock's theorem?

We could consider more than 4-dimensions. However, so far we experimentally detect only 4. Moreover, as long as we are interested in low energies and a phenomenological description, one is justified to expect that for any higher-dimensional theory there exist a 4-dimensional effective theory. If this theory is not general relativity, then it will have to contradict one of the other three assumptions. Going beyond the 4-dimensional effective description will be necessary in order to explain various characteristics of the theory which might seems ad hoc or unnatural when one is judging naively based on the 4-dimensional picture ({\em e.g.}~why the action has a certain form or why some couplings have specific values). But the 4-dimensional effective description should usually be adequate to discuss low-energy phenomenology and viability.

If we were to allow the equation of motion to be higher than second order partial differential equations (PDEs), then we would be generically introducing more degrees of freedom. This can be intuitively understood by considering the initial data one would have to provide when setting up an initial value problem in this theory (assuming that an initial value problem would be well posed). For instance, consider for simplicity a 4th order ordinary differential equation: to uniquely determine the evolution one would need to provide the first 3 time derivatives as initial data. So, a theory with higher order equation will generically have more propagating modes. Increasing the differential order is actually quite unappealing, as it leads to serious mathematical complications --- higher-order PDEs are not easy to deal with --- and serious stability issue. These will be discussed shortly.

Finally, one could give up diffeomorphism invariance. However, it has been long known that symmetries can be restored by introducing extra fields. This procedure  is known as the Stueckelberg mechanism, see Ref. \cite{Ruegg:2003ps} for a review. In Stueckelberg's work the new field was a scalar field introduced to restore gauge invariance in a massive Abelian gauge theory. By choosing the appropriate gauge one does away with the Stueckelberg field (it becomes trivial) but the theory is no longer manifestly gauge invariant. The Stueckelberg mechanism can be generalised to other symmetries, and specifically to diffeomorphism invariance.\footnote{Erich Kretschmann argued in 1917 that any theory can be put in a generally covariant form, which led to a famous debate with Einstein. A covariant version of Newtonian gravity can be found in Ref.~\cite{MTW}.} Hence, one can choose to think of theories that are not invariant under diffeomorphisms as diffeomorphism-invariant theories with extra Stueckelberg fields. 

In the previous section we demonstrated that diffeomorphism invariance has a central role in relating energy conservation and geodesic motion to the requirement that matter fields are on shell. From this discussion it also follows that if Stueckelberg fields are required in order to write a theory in a manifestly diffeomorphism invariant formulation, then these fields should not appear in the matter action, as is the case for any field that coupled non-minimally to gravity.

To summarised, we have argued that irrespectively of which of the 4 assumption of Lovelock's theorem one choses to relax, the outcome is always the same: one ends up with more degrees of freedom. The name of the game in alternative theories of gravity is, therefore, to tame the behaviour of these degrees of freedom. 

Clearly, many of the statements made in this section where rather heuristics and we relied heavily on the reader's intuition. In Section \ref{sec:3} convincing examples from scalar-tensor gravity that demonstrate all of the above will be presented.

\subsection{Taming the extra degrees of freedom}

Consider a simple system of two harmonic oscillators, describe by the lagrangian
\begin{equation}
L=\frac{1}{2}\dot{q}_1^2-\frac{1}{2}q_1^2+\frac{1}{2}\ddot{q}_2^2-\frac{1}{2} q_2^2\,.
\end{equation}
If I were to flip the sign of $q_1^2$ in the lagrangian $q_1$ would have exhibit exponential growth. If instead, I were to flip the sign of $\dot{q}_1^2$, the corresponding hamiltonian would not be bound from below. Having the wrong sign in front of certain terms renders the system unstable, but luckily in simple systems such as harmonic oscillators it is easy to know which sign to choose. In fact, coupling the two oscillators minimally would not affect this choice. 
Things become significantly more complicated though when one has degrees of freedom that couple non-minimally. Imagine adding a term such as $q_1^2 q_2^2$, $f(q_1) \dot{q}_2^2$ or $\dot{q}_1\dot {q}_2$. It is no longer obvious whether you system is stable or not. 

The situation is no different in a field theory. Fields whose hamiltonian is not bound from below are called ghosts and sensible theories are expected to be free of them. At the perturbative level this means that excitation around a certain configuration should have the right sign in front of the kinetic term. One also expects that physical configurations are classically stable, {i.e.}~all excitation around them have real propagation speeds. A complication that is always present in alternative theories of gravity is that the extra degrees of freedom are always non-minimally coupled to gravity (else there would be matter fields by definition). So, when constructing an action for a theory with a given field content it is nontrivial to judge whether it will satisfy the stability criteria mentioned above. As a result, one of the first calculations one does in every alternative theory of gravity is to check if all excitations satisfy these criteria around flat space (or some maximally symmetric space --- the vacuum solution of the theory).

In section \ref{lessassum} we mentioned that theories that lead to higher-order equation are generically plagued by instabilities. These instabilities are essentially due to the presence of ghosts. It has been shown by Ostrogradski in 1850 that non-degenerate Lagrangians with higher-order derivatives generically lead to Hamiltonians that are linear in at least on of the momenta \cite{ostro}. Such Hamiltonians are not bound from below. A detailed discussion can be found in Ref.~\cite{Woodard:2006nt}. Obviously, Ostragradski's instabilities make higher-order theories particularly unappealing. However, higher-order theories which can be explicitly re-written as second-order theories with more fields evade such instabilities. We will see an example of such a theory below.

Once stability issues have been addressed, and the behaviour of the new degrees of freedom has been tamed, the next step is to find a mechanism that hides them in regimes where general relativity is well tested and no extra degrees of freedom have been seen, but still allows them to be present and lead to different phenomenology in other regimes. How challenging a task this is and how inventive we have been in order to circumvent the difficulties will be demonstrated by the examples from scalar-tensor gravity presented in the next section.

It should be mentioned that a road less taken is to consider alternative theories with non-dynamical extra degrees of freedom. In fact, one could circumvent Lovelock's theorem by considering a gravity theory where fields other than the metric are present, but they are auxiliary field, so they do not satisfy dynamical equation but can be instead algebraically eliminated. This way ones has the same degrees of freedom as in general relativity and does not have to worry about instabilities associated with new dynamical fields. However, such an approach is not without serious shortcomings, see Ref.~\cite{Pani:2013qfa} for a discussion and references therein. For the rest of these notes we will focus one theories with dynamical new degrees of freedom, as most popular alternative theories of gravity fall under this category.

\section{Scalar-tensor gravity}
\label{sec:3}

\subsection{The prototype: Brans--Dicke theory}
\label{BD}

The action for Brans--Dicke theory is
\begin{eqnarray}
\label{bdaction}
S_{\rm BD}=\frac{1}{16\pi G}\int d^4x \sqrt{-g} \left(\varphi R-\frac{\omega_0}{\varphi} \nabla^\mu \varphi \nabla_\mu \varphi- V(\varphi)\right)+S_m(g_{\mu\nu},\psi)\,,
\end{eqnarray}
where $\varphi$ is a scalar field and $\omega_0$ is know as the Brans--Dicke parameter. After some manipulations, the corresponding field equation can take the form 
\begin{eqnarray}
R_{\mu\nu}-\frac{1}{2} R g_{\mu\nu}& = & \frac{8\pi G}{\varphi} T_{\mu\nu}+
\frac{\omega_0}{\varphi^2}\left(\nabla_\mu\varphi\nabla_\nu\varphi 
-\frac{1}{2}g_{\mu\nu} \, \nabla^\lambda\varphi\nabla_\lambda\varphi\right) \nonumber\\
&&\nonumber\\
& &+ \frac{1}{\varphi}\left(\nabla_\mu\nabla_\nu\varphi 
-g_{\mu\nu}\Box\varphi\right)-\frac{V(\varphi)}{2\varphi} 
g_{\mu\nu}\,,\\
\label{bdscalar}
(2\omega_0+3)\Box\varphi & = &\varphi \, V'-2V +8\pi G\, T \,,
\end{eqnarray}
where  
$\Box=\nabla^\lambda\nabla_\lambda$ and a 
prime denotes differentiation with respect to the argument.
In its original formulation Brans--Dicke theory did not have a potential. 

It is straightforward to see that in vacuo, where $T_{\mu\nu}=0$, the theory admits solutions where with $\varphi=\varphi_0=$constant, provided that $\varphi_0 \, V'(\varphi_0)-2V(\varphi_0)=0$. For such solution the metric actually satisfies Einstein's equations with an effective cosmological constant $V(\phi_0)$. So, one could be misled to think that, as long as $V(\varphi_0)$ has the right value, the predictions of the theory could be the same as those of general relativity. For instant, the space-time around the Sun could be described by such a solution, and then solar system constraints would be automatically satisfied. What invalidates this logic is that the $\varphi=\varphi_0$ solutions are not unique. $\varphi$ could actually have a nontrivial configuration, which would also force the metric to deviate for the corresponding solution of general relativity. 

This is indeed the case for spherically symmetric solution that describe the exterior of stars, and in particular the Sun. Consider for concreteness the case where $V=m^2(\varphi-\varphi_0)^2$. performing a newtonian expansion one can calculate the newtonian limit of the metric. The perturbations of the metric are
\begin{eqnarray}
h_{00}&=&\frac{GM_s}{\varphi_0 r}\left(1-\frac{1}{2\omega_0+3}exp\left[-\sqrt{\frac{2\varphi_0}{2\omega_0+3}}mr\right]\right)\,,\\
h_{ij}&=&\frac{GM_s}{\varphi_0 r}\delta_{ij}\left(1+\frac{1}{2\omega_0+3}exp\left[-\sqrt{\frac{2\varphi_0}{2\omega_0+3}}mr\right]\right)\,,
\end{eqnarray}
where $M_s$ is the mass of the Sun. There is a Yukawa-like correction to the standard $1/r$ potential, with effective mass $m_{\rm eff}=\sqrt{\frac{2\varphi_0}{2\omega_0+3}}m$ and range $m_{\rm eff}^{-1}$.
The ratio of the perturbations of the time-time component $h_{00}$ over any space-space diagonal component $h_{ij}|_{i=j}$, which is also known as the $\gamma$ (Eddington) parameter is then given by \cite{Perivolaropoulos:2009ak}
\begin{equation}
\gamma\equiv\frac{h_{ij}|_{i=j}}{h_{00}}=\frac{2\omega_0+3-exp\left[-\sqrt{\frac{2\varphi_0}{2\omega_0+3}}mr\right]}{2\omega_0+3+exp\left[-\sqrt{\frac{2\varphi_0}{2\omega_0+3}}mr\right]}\,.
\end{equation}

It is clear that in order for $\gamma$ to be close to 1, which is the value it has in general relativity, either $\omega_0$ or $m_{\rm eff}$ should be very large. Indeed, in the limit where $\omega_0\to \infty$ or $m\to \infty$ the equation imply that $\varphi\to \varphi_0$ and the constant $\varphi$ solutions with $g_{\mu\nu}$ satisfying Einstein's equation become unique. Current constraints on $\gamma$ require that $\gamma-1=(2.1\pm 2.3)\times 10^{-5}$ \cite{Bertotti:2003rm}. For $m=0$, this constraint would require $\omega_0$ to be larger than $40000$, which would make the theory indistinguishable from general relativity at all scales. For $\omega_0=O(1)$, the range of the Yukawa correction would have to be below the smaller scale we have currently tested the inverse square law, {\em i.e.}~a few microns. But if this is indeed the case, then this correction will never play a role at large scales.

The main message here is that weak gravity constraints are very powerful. It seems very hard to satisfy them and still have a theory whose phenomenology differs from that of general relativity at scales where we currently test gravity. One would have to circumvent this problem in order to construct a theory which is phenomenologically interesting.

\subsection{Scalar-tensor theories}

Scalar-tensor theories are straightforward generalisations of Brans-Dicke theory in which $\omega_0$ is promoted to a general function of $\varphi$. Their action is
\begin{eqnarray}
\label{staction}
S_{\rm st}&=&\frac{1}{16\pi G}\int d^4x \sqrt{-g} \Big(\varphi 
R-\frac{\omega(\varphi)}{\varphi} \nabla^\mu \varphi 
\nabla_\mu \varphi-V(\varphi) \Big)
+S_m(g_{\mu\nu},\psi)\,.
\end{eqnarray}
This is the most general action one can write for a scalar field non-minimally coupled to gravity which is second order in derivatives of the scalar. It can, therefore, be thought of as an effective field theory which captures, at some appropriate limit, the phenomenology of a more fundamental theory that contains a scalar field. 
The corresponding field equations are, after some manipulations
\begin{eqnarray}
R_{\mu\nu}-\frac{1}{2} R g_{\mu\nu}& = &
\frac{8\pi G}{\varphi} T_{\mu\nu}+ \frac{\omega(\varphi)}{\varphi^2}\left(\nabla_\mu\varphi\nabla_\nu\varphi 
-\frac{1}{2}g_{\mu\nu} \, \nabla^\lambda\varphi\nabla_\lambda\varphi\right) \nonumber\\
&&\nonumber\\
\!\!\!& + &\!\!\!\frac{1}{\varphi}\left(\nabla_\mu\nabla_\nu\varphi 
-g_{\mu\nu}\Box\varphi\right)-\frac{V(\varphi)}{2\varphi} 
g_{\mu\nu}\,,\\
\label{jfphi}
\left[2\omega(\varphi)+3\right]\Box\varphi & = &-\omega'(\varphi)
\, \nabla^\lambda\varphi \nabla_\lambda\varphi+\varphi \, V'-2V +8\pi G\, T  \,.
\end{eqnarray}

Scalar-tensor theories have been extensively studied and we will not review them here. See Refs.~\cite{fujiibook, faraonibook} for detailed reviews. The behaviour of the theories in the weak field limit will be no different than that of Brans--Dicke theory, though allowing $\omega$ to be a function of $\varphi$ will lead to a novel way of getting exciting phenomenology in the strong gravity regime, as we will see shortly.

We have given the action and field equations of scalar-tensor theory in terms of the metric that minimally couples to matter, $g_{\mu\nu}$. This is referred to as the Jordan frame. It is fairly common to re-write them in a different conformal frame, know as the Einstein frame, in which the (redefined) scalar couples minimally to gravity but it also couples to the matter. 

The 
conformal  transformation $\hat{g}_{\mu\nu}=\varphi 
\, g_{\mu\nu}$, together with the scalar field redefinition
%\begin{equation}
$4\sqrt{\pi}\varphi d\phi=\sqrt{2\omega(\varphi)+3} \, d\varphi$,
%\end{equation}
brings the action~(\ref{staction}) to the form
\begin{equation}
\label{stactionein}
S_{\rm st}=\int d^4x \sqrt{-\hat{g}} \Big(\frac{\hat{R}}{16\pi}-\frac{1}{2} \hat{g}^{\nu\mu}\partial_\nu \phi \partial_\mu \phi-U(\phi)\Big)+S_m(g_{\mu\nu},\psi)\,,
\end{equation}
where  $U(\phi)=V(\varphi)/\varphi^2$, $\hat{g}_{\mu\nu}$ is Einstein frame metric and all quantities with a hat are defined with this metric.
The field equations in the Einstein frame take the form
\begin{eqnarray}
\label{field1}
\hat{R}_{\mu\nu}-\frac{1}{2} \hat{R} \hat{g}_{\mu\nu}&=&8\pi G\,T_{\mu\nu}^\phi+\frac{8\pi G}{\varphi(\phi)}T_{\mu\nu}\,,\\
\label{field2}
\hat{\Box} \phi- U'(\phi)&=&\sqrt{\frac{4\pi G}{(2\omega+3)}}T\,,
\end{eqnarray}
 where
\begin{equation}
T_{\mu\nu}^\phi=\nabla_\mu \phi \nabla_\nu \phi 
-\frac{1}{2}g_{\mu\nu}\nabla_\lambda \phi \nabla^\lambda \phi 
-U(\phi)g_{\mu\nu}\,,
\end{equation}
whereas $T_{\mu\nu}$ and $T$ are the Jordan frame stress-energy tensor and its trace respectively.

The fact that $\phi$ couples minimally to $\hat{g}_{\mu\nu}$ in the Einstein frame makes calculations much simpler in many cases, especially in vacuo, where the theory becomes general relativity with a minimally coupled scalar field. 
One can use any of the two frames to perform calculations but some care is needed when interpreting results that do not involve conformally invariant quantities. The physical significance of the two metrics, $g_{\mu\nu}$ and $\hat{g}_{\mu\nu}$, should be clear: the former is the metric whose geodesics will coincide with test particle trajectories, as it couples minimally to matter. The latter, is just a special choice which brings the action in a convenient form. See Ref.~\cite{Sotiriou:2007zu} and references therein for more detailed discussions.

\subsection{Hiding the scalar field, part I}

\label{hiding1}

We will now briefly discuss some mechanisms that can hide the scalar field in the weak field regime near matter but still allow the theory to deviate significantly from general relativity in cosmology or in the strong gravity regime. 

The first and oldest of these mechanisms is present in theories were $\omega(\varphi)$ diverges for some constant value of $\varphi$ \cite{Damour:1993hw,Damour:1996ke}. Consider the case where there is no potential. In configurations where $\omega\to\infty$ one essentially ends up with a constant scalar and metrics that satisfy Einstein's equations. This follows intuitively by the analysis of the newtonian limit of Brans--Dicke theory when $\omega_0\to \infty$, or more rigorously by inspecting the field equations or the action. It is more convenient and straightforward to consider the Einstein frame. In the absence of a potential, Eq.~(\ref{field2}) admits $\phi=\phi_0=$constant solutions with $\omega(\phi_0)\to \infty$ even inside matter.\footnote{If there is a potential $\phi=\phi_0$ solutions are only admissible if $U'(\phi_0)=0$ as well.} For such solutions Eqs.~(\ref{field1}) reduce to Einstein's equations (with a rescaled c`oupling inside matter). Going back to the Jordan frame, such solutions correspond to $\varphi=$constant with $g_{\mu\nu}$ satisfying Einstein's equations.

A key difference with Brans--Dicke theory with very large $\omega_0$ is that here $\omega$ diverges only in the specific configuration for the scalar, so one needs to check under which circumstances such configurations are solution of the physical system of interest. In other words, one has to check that $\varphi$ will be dynamically driven into this configuration in situations where one would like to recover general relativity.

It has been indeed shown in Refs.~\cite{Damour:1993hw,Damour:1996ke} that there exist theories where in principle both $\phi=\phi_0$ and non-trivial $\phi$ solutions exist for stars. Which of the two configurations will be realised after gravitational collapse depends (roughly speaking) on the compactness of the star. For ordinary stars, such as the Sun, the constant scalar solution is the one realised. The metric describing their exterior is then the same as in general relativity and this makes the theories indistinguishable from the latter in the Solar system. For compact stars instead, such as neutron stars, the non-trivial scalar configuration becomes energetically favourable and the metric significantly deviates from the one general relativity would yield. Hence, the strong-field phenomenology will be distinct from that of general relativity. The importance of this result lies on the fact that it was the first demonstration that one can construct a theory which agrees with general relativity in the weak field limit but still gives distinct and testable predictions in the strong field regime. There is a very sharp transition from the $\phi=$constant to the non-trivial $\phi$ configurations as one increases the compactness of the start, so the mechanism that causes this transition is called ``spontaneous scalarization'' \cite{Damour:1993hw,Damour:1996ke}.

This mechanism relies entirely on the functional form of $\omega$, which turned out to be intimately related to how the scalar field is sourced by matter. There is a different type of mechanism to hide the scalar field that relies on the potential $V$, or $U$, and is called the chameleon mechanism \cite{Khoury:2003aq}. In terms of the newtonian limit of Brans--Dicke theory that was given in section \ref{BD} the chameleon mechanism can be thought of as a dependence of the effective mass, and the corresponding range of the Yukawa-like correction, on the characteristics of a given matter configuration. As discussed earlier, when the effective mass gets large enough, the range of the Yukawa-like correction becomes short enough to be negligible in any known experiment. But if one wants the scalar field to have any effect in cosmology, for example to account for dark energy, then the range of the correction should actually be long. The dependence of the mass on the nearby matter configuration makes it possible to have it both ways.

For a scalar field that experiences only self interactions one defines as the mass the value of the second derivative of its potential at the minimum of the potential. However, things are slightly more complicated for non-minimally coupled scalar fields. It is easier to resort to the Einstein frame and consider eq.~(\ref{field2}). Then $\phi$'s dynamics are governed by en effective potential $U_{\rm eft}=U(\phi)+(\ln \varphi) T/2$ [as $U'_{\rm eft}=U'(\phi)+\sqrt{4\pi G/(2\omega+3)}T$]. By choosing $U$ appropriately (the behaviour of $\omega$ is much less relevant) one can arrange that $\phi$ have a very small mass when $T$ is small and a very large mass when $T$ is large, as the term $(\ln \varphi) T/2$ clearly deforms the potential. The most characteristic example is when choosing $U\sim e^{-\phi}$ and $\omega$ is a constant, so that the $T$-dependent deformation is linear in $\phi$. Without this deformation the range of the force would be infinite. But the deformations introduces a minimum that leads to a short range force.

There are two subtleties in the line of reasoning we just laid out, which are sometimes not given enough attention in the literature. Firstly, we used the Einstein frame, but the mass that determines the range of the Yukawa-like correction is not actually the one associated with the effective potential of $\phi$ in this frame (neither the one defined as $V''(\varphi_0)$ in the Jordan frame actually, hence the use of $m_{\rm eff}$ in section \ref{BD}). However, one can show that the various masses are intimately related \cite{Faraoni:2009km}. Secondly, Solar system test are not really performed in a high density environment but in vacuo, outside a high density matter configuration. On the other hand, continuity of the scalar field profile implies that, even outside the star, there will be a region for which the configuration will be influenced more by the interior configuration through boundary conditions that by the asymptotic configuration. We refer the reader to a recent review on the chameleon mechanism for a thorough discussion \cite{Khoury:2013yya}.

A third mechanism for hiding the scalar field in the Solar system is the symmetron mechanism \cite{Hinterbichler:2010es}. Here both the form of $\omega$ and the form of the potential are important. In the Einstein frame the potential $U$ is assumed to have the form
\begin{equation}
U(\phi)=-\frac{1}{2}\mu^2 \phi^2+\frac{1}{4} \lambda\phi^4\,.
\end{equation}
In the absence of matter $\phi$ would then have a minimum at $\phi_0=\mu/\sqrt{\lambda}$. The value of the potential at the minimum is related to an effective cosmological constant, which one can tune to the desired value by appropriately choosing $\mu$ and $\lambda$. Assume now that $\omega$ has such a functional dependence on $\varphi$ (and implicitly on $\phi$) that in the presence of matter the effective potential would be
\begin{eqnarray}
U_{\rm eff}(\phi)&=&-\frac{1}{2}\mu^2 \phi^2+\frac{1}{4} \lambda\phi^4+(1+\frac{\phi^2}{M^2}) \frac{T}{2}\nonumber\\
&=&\frac{1}{2}\left(\frac{T}{M^2}-\mu^2\right)\phi^2+\frac{1}{4} \lambda\phi^4+\frac{T}{2}\,,
\end{eqnarray}
where $M$ is a characteristic mass scale, and 
\begin{equation}
U'_{\rm eff}(\phi)=-\mu^2 \phi+ \lambda\phi^3+\frac{\phi}{M^2}T\,,
\end{equation}
For such a choice, $\omega(\phi=0)\to \infty$. Provided that $T/M^2>\mu^2$, $\phi=0$ becomes the minimum of the effective potential and eq.~(\ref{field2}) admits $\phi=0$ solution in the presence of matter. 

In a certain sense, there is some similarity between the symmetron mechanism and the models that exhibit spontaneous scalarization in compact stars discussed earlier. In fact, one could see the symmetron mechanism as a cosmological scalarization. The way the symmetron mechanism works in a realistic matter configuration is actually more complicated than the simplistic description given above. For example, in a realistic matter configuration, the scalar has to smoothly change from being zero inside the matter to obtaining its non-zero asymptotic value outside the matter. We refer the reader to Ref.~\cite{Hinterbichler:2010es} for more details.

\subsection{The Horndeski Action}

The action of scalar-tensor theory in eq.~(\ref{staction}) is the most general action that is quadratic in derivatives of the scalar, up to boundary terms. It is not, however, the most general action that can lead to second order field equations for the metric and the scalar. Horndeski has shown that the most general action with this property is \cite{Horndeski:1974wa}
\begin{eqnarray}
\label{hdaction}
S_H&=&\int d^4 x \sqrt{-g}\left(L_2+L_3+L_4+L_5\right)\,,
\end{eqnarray}
where
\begin{align}
L_2 &= K(\phi,X)     ,
\\
L_3 &= -G_3(\phi,X) \Box \phi     ,
\\
L_4 &= G_4(\phi,X) R + G_{4X} \left[ (\Box \phi)^2 
-(\nabla_\mu\nabla_\nu\phi)^2 \right]     ,
\\
L_5 &= G_5(\phi,X) G_{\mu\nu}\nabla^\mu \nabla^\nu \phi - 
 \frac{G_{5X}}{6} \left[ (\Box \phi)^3 - 3\Box 
\phi(\nabla_\mu\nabla_\nu\phi)^2 + 2(\nabla_\mu\nabla_\nu\phi)^3 \right] \,,
\end{align}
the $G_i$ are unspecified functions of $\phi$ and $X\equiv-\frac{1}{2} \nabla^\mu \phi \nabla_\mu \phi$ and  $G_{iX}\equiv \partial G_i/\partial X$. Scalar fields described by this action are also known as Generalised Galileons \cite{Deffayet:2009mn}. The name comes from a particular class of scalar theories in flat space which enjoy Galilean symmetry, {\em i.e.}~symmetry under $\phi \to \phi + c_\mu x^\mu +c$, where $c_\mu$ is a constant one-form and $c$ is a constant \cite{Nicolis:2008in}. These fields are known as Galileons. A certain subclass of Generalised Galileons reduce to Galileons in flat space. But galilean symmetry itself does not survive the passage to curved space \cite{Deffayet:2009wt} (it is local symmetry) and the full Horndeski action does not reduce to the Galileon action in flat space.\footnote{The numbering of the terms in the Lagrangian, $L_2$ to $L_5$, is also a remnant of the original flat space Galileons \cite{Nicolis:2008in}. The index indicates there the number of copies of the field in each term. In the Generalised Galileons the $L_i$ term contains $i-2$ second derivatives of the scalar.}

Horndeski's theory is intrinsically interesting as a field theory, as it contains more than two derivatives in the action but still leads to second order equations. That comes at the price of having highly nonlinear derivative (self-)interactions. It is worth noting that, even though Horndeski's actions includes second derivatives of the fields, it avoids Ostrogradski's instability because it does not satisfy the non-degeneracy assumption.\footnote{The Einstein--Hilbert action also contains second derivatives of the metric and is degenerate, thus avoiding Ostrogradski's instability.}

A more detailed discussion about the characteristics of the theory goes beyond the scope of these lecture notes, so we refers the reader to Ref.~\cite{Deffayet:2013lga} for a recent review.

\subsection{Hiding the scalar field, part II}

The high degree of non-linearity in the scalar field equations of Hordenski's theory certainly makes them mathematically complicated. However, it does not come without advantages. In regimes where these highly non-linear terms will dominate over the standard Brans--Dicke-like terms the behaviour of the scalar field will be significantly different from that of the Brans--Dicke scalar discussed above. In fact, such theories can exhibit the ``Vainshtein effect'': solutions of the linearised version of the theory --- in which the higher derivative terms would give no significant contribution --- can be very different from solutions of general relativity, but fully non-linear solutions might be indistinguishable from those of the latter. The term ``Vainshtein effect" originates from massive gravity theory where the mechanism was first demonstrated by Vainshtein in Ref.~\cite{Vainshtein:1972sx}. A detailed introduction to the Vainshtein mechanism can be found in Ref.~\cite{Babichev:2013usa}.

\section{Scalar-tensor gravity in disquise}
\label{sec:4}

In section \ref{lessassum} it was argued that allowing for higher-order field equations or giving up diffeomorphism invariance leads to more degrees of freedom. In this section we provide two examples that support this claim. In both cases the new degree of freedom is a scalar field and this can be made explicit, either by field redefinitions, or via the Stueckelberg mechanism.

\subsection{$f(R)$ gravity}

The action of $f(R)$ gravity is
\begin{equation}
S=\frac{1}{16\pi G}\int d^4 x \sqrt{-g} f(R) +S_m(g_{\mu\nu},\psi)\,,
\end{equation}
where $f$ is some function of the Ricci scalar of $g_{\mu\nu}$. Variation with respect to the metric $g^{\mu\nu}$ yields
\begin{equation}
f'(R) R_{\mu\nu}-\frac{1}{2} f(R) g_{\mu\nu} -[\nabla_\mu \nabla\nu- g_{\mu\nu} \Box] f'(R)=8\pi G T_{\mu\nu}\,.
\end{equation}
Provide that $f''(R)\neq 0$, in which case the theory would be general relativity, these are clearly 4th-order equations in $g_{\mu\nu}$. One would then expect the theory to suffer from the Ostrogradski instability mentioned earlier. 

Consider now the action
\begin{equation}
S=\frac{1}{16\pi G}\int d^4 x \sqrt{-g} \left[f(\phi)+\varphi(R-\phi) \right]+S_m(g_{\mu\nu},\psi)\,.
\end{equation}
Variation with respect to $\varphi$ yields $\phi=R$. Replacing this algebraic constraint back in the action yields the action of $f(R)$ gravity. Hence, the two actions are (classically) dynamically equivalent. If instead one varies with respect to $\phi$ one gets $\varphi=f'(\phi)$. Replacing this algebraic relation back in the action one gets another dynamically equivalent action
\begin{equation}
S=\frac{1}{16\pi G}\int d^4 x \sqrt{-g} \left[\varphi R-V(\varphi) \right]+S_m(g_{\mu\nu},\psi)\,,
\end{equation}
where $V(\varphi)\equiv f(\phi)-\phi f'(\phi)$ ($V$ is essentially the Legendre transform of $f$). This theory is actually a Brans--Dicke theory with vanishing $\omega_0$, also known as the O'Hanlon action \cite{O'Hanlon:1972ya}.

This simple exercise establishes that $f(R)$ gravity can be recast in the form of a special Brans--Dicke theory, something that has been know for quite a while, see {\em e.g.}~Ref.~\cite{Teyssandier:1983zz}.
It demonstrates both how higher-order theories propagate more degrees freedom --- in this case a scalar --- and how such theories avoid Ostrogradski's instability when they can be recast into second-order theories with more degrees of freedom.

\subsection{Ho\v rava gravity}

Ho\v rava gravity \cite{arXiv:0901.3775} is a theory with a preferred spacetime foliation. The action of the theory is \cite{Blas:2009qj}
\begin{equation}
\label{SBPSHfull}
S_{H}= \frac{1}{16\pi G_{H}}\int dT d^3x \, N\sqrt{h}\left(L_2+\frac{1}{M_\star^2}L_4+\frac{1}{M_\star^4}L_6\right)\,,
\end{equation}
where 
\begin{equation}
\label{L2}
L_2
=K_{ij}K^{ij} - \lambda K^2 
+ \xi {}^{(3)}\!R + \eta a_ia^i,
\end{equation}
where $T$ is the preferred time, $K_{ij}$ is the extrinsic curvature of the surfaces of the foliation and $K$ its trace, ${}^{(3)}\!R$ is the intrinsic curvature of these surfaces, $N$ is the lapse function, $h_{ij}$ is the induced metric and $h$ is the determinant of the induced metric, $a_i\equiv \partial_i \ln N$, $G_H$ is a coupling constant with dimensions of length squared and $\lambda$, $\xi$, and $\eta$ are dimensionless couplings. Since the action is written in a preferred foliation the theory does not enjoy invariance under diffeomorphisms. It is still invariant under the subset of diffeomoprhisms that respect the foliation, $T\to T'=f(T)$ and $x^i\to x'^i=x'^i(T,x^i)$. $L_4$ and $L_6$ include all possible terms that respect this symmetry and contain up to four and six spatial derivatives respectively. $M_\star$ is a characteristic mass scale suppressing these higher order terms. 

Ho\v rava gravity has been proposed as a power-counting renormalizable gravity theory and the presence of the higher-order terms in $L_4$ and $L_6$ is crucial in order to have the right UV behaviour \cite{arXiv:0901.3775}. However, these terms will not concern us here, as we intend to consider the low energy part of the theory, $L_2$, as an example of a gravity theory that does not respect diffeomorphism invariance. For a brief review on the basic features of Ho\v rava gravity see Ref.~\cite{Sotiriou:2010wn}.

Consider now the action
\begin{equation} \label{S}
S'= \frac{1}{16\pi G'}\int \sqrt{-g} (-R -M^{\alpha\beta}{}_{\mu\nu} \nabla_\alpha u^\mu \nabla_\beta u^\nu)
d^{4}x \end{equation}
where 
\begin{equation} M^{\alpha\beta}{}_{\mu\nu} = c_1 g^{\alpha\beta}g_{\mu\nu}+c_2\delta^{\alpha}_{\mu}\delta^{\beta}_{\nu}
+c_3 \delta^{\alpha}_{\nu}\delta^{\beta}_{\mu}+c_4 u^\alpha u^\beta g_{\mu\nu}\,,
\end{equation}
$c_i$ are dimensionless coupling constants and $u_\mu$ is given by 
\begin{equation}
\label{ho}
u_\mu=\frac{\partial_\mu T}{\sqrt{g^{\lambda\nu}\partial_\lambda T \partial_\nu T}}\,.
\end{equation}
This is a scalar-tensor theory where the scalar field $T$ only appears in the action in the specific combination of eq.~(\ref{ho}). Therefore, $u^\mu$ can be thought of as a hypersurface orthogonal, unit, timelike vector (as $u^\mu u_\mu=1$). The theory can be thought of as a restricted version of Einstein-aether theory \cite{Jacobson:2000xp,Jacobson:2008aj} where the aether is forced to be hyper surface orthogonal before the variation. 

Now, following the lines of Ref.~\cite{Jacobson:2010mx}, one can observe that $T$ always has a timelike gradient, so it can be a good time coordinate for any solution. Then one can give up some of the gauge freedom in order to re-write the theory in terms of this time coordinate. This involves introducing a foliation of $T=$constant hyper surfaces, to which $u^\mu$ will be normal, and re-writing the action in this foliation. Then $u_\mu=N \delta^0_\mu$, where $N$ is the lapse of this foliation, and action (\ref{S}) takes the form 
\begin{equation}
\label{SBPSHIR}
S'= \frac{1}{16\pi G_{H}}\int dT d^3x \, N\sqrt{h}\left(K_{ij}K^{ij} - \lambda K^2 + \xi {}^{(3)}\!R + \eta a_ia^i, \right)\,,
\end{equation}
where
\begin{equation}
\label{HLpar}
\frac{G_H}{G'}=\xi=\frac{1}{1-(c_{1}+c_3)}, \quad \lambda=\frac{1+c_2}{1-(c_{1}+c_{3})},\quad \eta=\frac{c_1+c_4}{1-(c_1+c_3)}\,.
\end{equation}
Action (\ref{SBPSHIR}) is clearly the infrared ($L_2$) part of action (\ref{SBPSHfull}), which means that the initial action (\ref{S}) is just the diffeomophism invariant version of the infrared limit of Ho\v rava gravity. $T$ can then be thought of as the Stueckelberg field one needs to introduce in order to restore full diffeomorphism invariance in Ho\v rava gravity. It is clearly a dynamical field and in the covariant picture one can think of it as having a nontrivial configuration which defines the preferred foliation in every solution. When the theory is written in the preferred foliation, as in eq.~(\ref{SBPSHfull}) then the scalar degree of freedom is no longer explicit, but one can expect its existence because the action has less symmetry.

\section{Scalar fields around black holes}
\label{sec:5}

As already mentioned in the introduction, black holes and compact stars are of particular interest in alternative theories of gravity as potential probes of the strong gravity regime. Black holes  in particular have the advantage of being vacuum solutions, so one need not worry about matter, and of containing horizons, hence they have a very interesting causal structure.

One could argue that the existence of extra degrees of freedom --- in this case a scalar field --- in a gravity theory will generically lead to black hole solutions that differ from their general relativity counterpart. They could then be used as probes for deviation from Einstein's theory, or even for the very existence of scalar fields. However, there are ``no-hair'' theorems is scalar-tensor gravity that suggest otherwise \cite{hawking2,Sotiriou:2011dz}. In particular, according to these theorems stationary, asymptotically flat black holes in the theories described by the action of eq.~(\ref{staction}) are identical to black holes in general relativity. This is because the scalar field is forced to have a $\phi=$constant configuration in stationary, asymptotically flat space times with a horizon. Quiescent astrophysical black holes that are the endpoints of gravitational collapse are stationary. They are also asymptotically flat to a very good approximation. Hence, one is tempted to believe that black holes in scalar-tensor theories will be indistinguishable from black holes in general relativity.

Such an interpretation of the no-hair theorems would be misleading for several reasons. First of all, a perturbed Kerr spacetime in a scalar-tensor theory would differ from a perturbed Kerr spacetime in general relativity, a characteristic example being the existence of a scalar mode in the gravitational wave spectrum \cite{Barausse:2008xv}. Secondly, cosmological asymptotics do induce scalar hair in principle \cite{Jacobson:1999vr}, though the deviation from the Kerr geometry is unlikely to be detectable \cite{Horbatsch:2011ye}. Finally, astrophysical black holes tend to be surrounded by matter in various forms --- companion stars, accretion disks, or the galaxy as a whole. Eq.~(\ref{jfphi}) or eq.~(\ref{field2}) imply that, in the presence of matter, constant scalar solutions are only allowed in theories for which $\omega$ diverges at the minimum of the potential. This has been already discussed in section \ref{hiding1} (theories that exhibit ``spontaneous scalarization'' \cite{Damour:1993hw,Damour:1996ke}). Hence, generically the presence of matter around the black hole will tend to induce scalar hair and the pending question is to determine how important this effect might be.

So, when put in astrophysical context, the no-hair theorems tell us that black holes that are endpoints of collapse will be rather close to the Kerr solution and that we can use perturbative techniques in order to study phenomena around them (which provides an important simplification). They do not, however, imply that astrophysical black holes in scalar-tensor gravity are indistinguishable from astrophysical black holes in general relativity. In fact, it has been suggested that there might be smoking gun effects associated with the scalar field in scalar-tensor theories. For example, in Ref.~\cite{Cardoso:2011xi} it has been shown that there exist floating orbits around Kerr black holes in these theories, i.e. a  particles can orbit the black holes without ``sinking'' into it even though gravitational radiation is emitted. The loss of energy of the emission is balanced by loss of angular momentum of the black hole. In Ref.~\cite{Cardoso:2013fwa} instead, it was shown that, in theories that admit a constant scalar configuration in the presence of matter, black holes can undergo spontaneous scalarization or exhibit instabilities related to superradiance and very large amplification factors for superradiant scattering.

We now move on to black holes in generalised scalar-tensor gravity, {\em i.e.}~theories described by the Horndeski action in eq.~(\ref{hdaction}). There are no no-hair theorems covering the complete class of theories. On the contrary, there are already black hole solutions that have non-trivial scalar field configurations in theories that belong to this class, see {\em e.g.}~Ref.~\cite{Kanti:1995vq}. It has been claimed in Ref.~\cite{Hui:2012qt} that in the subclass of theories in which the scalar enjoys shift symmetry, {\em i.e.}~symmetry under $\phi \to \phi +$ constant, only trivial scalar configuration are admissible for static, spherically symmetric and asymptotically flat black holes, and, hence, these black holes are described by the Schwarzschild solutions. It has been argued in Ref. [49] that, when valid, the no-hair theorem of Ref. [48] can straightforwardly be generalised to slowly rotating black holes. However, it has been also been shown there that the theorem holds in the first place only if one forbids a linear coupling between the scalar field and the Gauss--Bonnet invariant. Such a coupling is allowed by shift symmetry, since the Gauss--Bonnet term is a total diverge. A term that contains this coupling is implicitly part of the Horndeski action, even though the representation of eq.~(\ref{hdaction}) does not make that manifest. One can impose symmetry under $\phi \to -\phi$ in order to do away with this term (together with various others in the action). However, the conclusion is that the  subclass of theories for which one can have a no-hair theorem is more limited than originally claimed.

We close this section with a few remarks on black holes in Lorentz-violating theories, since, as we argued above Ho\v rava gravity can be re-written as a scalar-tensor theory. One could question whether black holes can actually exist in this theory, as well as in other Lorentz-violating theories, as one can have perturbations that travel with arbitrarily high speed and could, therefore, penetrate conventional horizons.\footnote{Ho\v rava gravity exhibits instantaneous propagation even at low energies \cite{Blas:2011ni}, and on general grounds one would expect the UV completion of any Lorentz violating theory to generically introduce higher order dispersion relations.} However, it has been shown that a new type of horizon that shields its exterior from any signal that comes from its interior, irrespectively of how fast it propagates, can exist in theories with a preferred foliation, called the universal horizon \cite{Blas:2011ni, Barausse:2011pu,Barausse:2012ny,Barausse:2012qh}. The existence of such a horizon implies that the notion of a black hole can exist in Lorentz-violating theories. For a thorough discussion on this topic see Ref.~\cite{Barausse:2013nwa}.

\section{Conclusions}
\label{conc}

In these lecture notes I have attempted to highlight some interesting concepts, pitfalls and subtleties that appear when one goes beyond general relativity. Perhaps it is helpful to list the most important ones:
\begin{itemize}
\item Any attempt to modify the action of general relativity will generically lead to extra degrees of freedom (carefully engineered exceptions can exist);
\item These degrees of freedom may be manifest as extra dynamical fields or may be implicit because of higher order equations or less symmetry;
\item The actual number of degrees of freedom might be quite obscure in some specific field representation;
\item Taming the behaviour of these extra degrees of freedom is what constructing viable and successful (in terms of some desirable phenomenological signature) models is about;
\item One should constantly be seeking for new constraints on deviations from general relativity, and the strong gravity regime is particularly promising in this respect.
\end{itemize}

A brief review of gravity theories  with an extra scalar degree of freedom has been given and some of their basic features have been discussed. Even though I touched upon virtually all such theories, these lecture notes do not constitute a thorough review of the theories and their phenomenology. I have simply selectively discussed specific aspects of each theory in an attempt to provide useful examples for the points listed above.

\begin{acknowledgement}
The research leading to these results has received funding from the European Research Council
under the European Union's Seventh Framework Programme (FP7/2007-2013) / ERC
Grant Agreement n. 306425 ``Challenging General Relativity".
\end{acknowledgement}
%

%%%%%%%%%%%%%%%%%%%%%%%% referenc.tex %%%%%%%%%%%%%%%%%%%%%%%%%%%%%%
% sample references
% %
% Use this file as a template for your own input.
%
%%%%%%%%%%%%%%%%%%%%%%%% Springer-Verlag %%%%%%%%%%%%%%%%%%%%%%%%%%
%
% BibTeX users please use
% \bibliographystyle{}
% \bibliography{}

\begin{thebibliography}{99.}

%\cite{AmelinoCamelia:2004hm}
\bibitem{AmelinoCamelia:2004hm} 
  G.~Amelino-Camelia,
  %``Introduction to quantum-gravity phenomenology,''
  Lect.\ Notes Phys.\  {\bf 669}, 59 (2005)
  [gr-qc/0412136].
  %%CITATION = GR-QC/0412136;%%
  %37 citations counted in INSPIRE as of 07 Apr 2014

%\cite{Liberati:2013xla}
\bibitem{Liberati:2013xla} 
  S.~Liberati,
  %``Tests of Lorentz invariance: a 2013 update,''
  Class.\ Quant.\ Grav.\  {\bf 30}, 133001 (2013)
  [arXiv:1304.5795 [gr-qc]].
  %%CITATION = ARXIV:1304.5795;%%
  %22 citations counted in INSPIRE as of 07 Apr 2014

%\cite{Ade:2013zuv}
\bibitem{Ade:2013zuv} 
  P.~A.~R.~Ade {\it et al.}  [Planck Collaboration],
  %``Planck 2013 results. XVI. Cosmological parameters,''
  arXiv:1303.5076 [astro-ph.CO].
  %%CITATION = ARXIV:1303.5076;%%
  %1327 citations counted in INSPIRE as of 27 Feb 2014
  
  %\cite{Weinberg:1988cp}
\bibitem{Weinberg:1988cp} 
  S.~Weinberg,
  %``The Cosmological Constant Problem,''
  Rev.\ Mod.\ Phys.\  {\bf 61}, 1 (1989).
  %%CITATION = RMPHA,61,1;%%
  %2514 citations counted in INSPIRE as of 27 Feb 2014
  
  %\cite{Carroll:2000fy}
\bibitem{Carroll:2000fy} 
  S.~M.~Carroll,
  %``The Cosmological constant,''
  Living Rev.\ Rel.\  {\bf 4}, 1 (2001).
 % [astro-ph/0004075].
  %%CITATION = ASTRO-PH/0004075;%%
  %930 citations counted in INSPIRE as of 27 Feb 2014


\bibitem{willbook} C.~Will, {\em Theory and experiment in gravitational physics}, (Cambridge University Press, Cambridge, 1993).

%\cite{Lovelock:1971yv}
\bibitem{Lovelock:1971yv} 
  D.~Lovelock,
  %``The Einstein tensor and its generalizations,''
  J.\ Math.\ Phys.\  {\bf 12}, 498 (1971).
  %%CITATION = JMAPA,12,498;%%
  %825 citations counted in INSPIRE as of 26 Feb 2014


%\cite{Lovelock:1972vz}
\bibitem{Lovelock:1972vz} 
  D.~Lovelock,
  %``The four-dimensionality of space and the einstein tensor,''
  J.\ Math.\ Phys.\  {\bf 13}, 874 (1972).
  %%CITATION = JMAPA,13,874;%%
  %103 citations counted in INSPIRE as of 26 Feb 2014
  
  %\cite{Ruegg:2003ps}
\bibitem{Ruegg:2003ps} 
  H.~Ruegg and M.~Ruiz-Altaba,
  %``The Stueckelberg field,''
  Int.\ J.\ Mod.\ Phys.\ A {\bf 19}, 3265 (2004).
%  [hep-th/0304245].
  %%CITATION = HEP-TH/0304245;%%
  %117 citations counted in INSPIRE as of 26 Feb 2014
  
  \bibitem{MTW} C.~W.~Misner, K.~S.~Thorne, J.~A.~Wheeler,  {\em Gravitation}, (W.~H.~Freeman, 1973).
  
    \bibitem{ostro} M.~Ostrogradski, Mem.\ Ac.\ St.\ Peterbourg {\bf VI 4}, 385 (1850).
  
  %\cite{Woodard:2006nt}
\bibitem{Woodard:2006nt} 
  R.~P.~Woodard,
  %``Avoiding dark energy with 1/r modifications of gravity,''
  Lect.\ Notes Phys.\  {\bf 720}, 403 (2007).
%  [astro-ph/0601672].
  %%CITATION = ASTRO-PH/0601672;%%
  %255 citations counted in INSPIRE as of 28 Feb 2014
  
  %\cite{Pani:2013qfa}
\bibitem{Pani:2013qfa} 
  P.~Pani, T.~P.~Sotiriou and D.~Vernieri,
  %``Gravity with Auxiliary Fields,''
  Phys.\ Rev.\ D {\bf 88}, 121502 (2013).
%  [arXiv:1306.1835 [gr-qc]].
  %%CITATION = ARXIV:1306.1835;%%
  %1 citations counted in INSPIRE as of 28 Feb 2014
  
  %\cite{Perivolaropoulos:2009ak}
\bibitem{Perivolaropoulos:2009ak} 
  L.~Perivolaropoulos,
  %``PPN Parameter gamma and Solar System Constraints of Massive Brans-Dicke Theories,''
  Phys.\ Rev.\ D {\bf 81}, 047501 (2010).
%  [arXiv:0911.3401 [gr-qc]].
  %%CITATION = ARXIV:0911.3401;%%
  %24 citations counted in INSPIRE as of 27 Feb 2014
  
  
  %\cite{Bertotti:2003rm}
\bibitem{Bertotti:2003rm} 
  B.~Bertotti, L.~Iess and P.~Tortora,
  %``A test of general relativity using radio links with the Cassini spacecraft,''
  Nature {\bf 425}, 374 (2003).
  %%CITATION = NATUA,425,374;%%
  %560 citations counted in INSPIRE as of 27 Feb 2014
  
  \bibitem{faraonibook} V.~Faraoni, {\em Cosmology in Scalar-Tensor Gravity}, (Springer, 2004).
  
  \bibitem{fujiibook} Y.~Fujii, K.~Maeda, {\em The Scalar-Tensor Theory of Gravitation}, (Cambridge University Press, 2003).
  
  %\cite{Sotiriou:2007zu}
\bibitem{Sotiriou:2007zu} 
  T.~P.~Sotiriou, V.~Faraoni and S.~Liberati,
  %``Theory of gravitation theories: A No-progress report,''
  Int.\ J.\ Mod.\ Phys.\ D {\bf 17}, 399 (2008).
%  [arXiv:0707.2748 [gr-qc]].
  %%CITATION = ARXIV:0707.2748;%%
  %37 citations counted in INSPIRE as of 28 Feb 2014
  
%\cite{Damour:1993hw}
\bibitem{Damour:1993hw} 
  T.~Damour and G.~Esposito-Farese,
  %``Nonperturbative strong field effects in tensor - scalar theories of gravitation,''
  Phys.\ Rev.\ Lett.\  {\bf 70}, 2220 (1993).
  %%CITATION = PRLTA,70,2220;%%
  %162 citations counted in INSPIRE as of 28 Feb 2014
  
  %\cite{Damour:1996ke}
\bibitem{Damour:1996ke} 
  T.~Damour and G.~Esposito-Farese,
  %``Tensor - scalar gravity and binary pulsar experiments,''
  Phys.\ Rev.\ D {\bf 54}, 1474 (1996).
%  [gr-qc/9602056].
  %%CITATION = GR-QC/9602056;%%
  %154 citations counted in INSPIRE as of 28 Feb 2014
  

%\cite{Khoury:2003aq}
\bibitem{Khoury:2003aq} 
  J.~Khoury and A.~Weltman,
  %``Chameleon fields: Awaiting surprises for tests of gravity in space,''
  Phys.\ Rev.\ Lett.\  {\bf 93}, 171104 (2004).
%  [astro-ph/0309300].
  %%CITATION = ASTRO-PH/0309300;%%
  %550 citations counted in INSPIRE as of 02 Mar 2014  

%\cite{Faraoni:2009km}
\bibitem{Faraoni:2009km} 
  V.~Faraoni,
  %``Scalar field mass in generalized gravity,''
  Class.\ Quant.\ Grav.\  {\bf 26}, 145014 (2009).
%  [arXiv:0906.1901 [gr-qc]].
  %%CITATION = ARXIV:0906.1901;%%
  %14 citations counted in INSPIRE as of 02 Mar 2014
  
  %\cite{Khoury:2013yya}
\bibitem{Khoury:2013yya} 
  J.~Khoury,
  %``Chameleon Field Theories,''
  Class.\ Quant.\ Grav.\  {\bf 30}, 214004 (2013).
%  [arXiv:1306.4326 [astro-ph.CO]].
  %%CITATION = ARXIV:1306.4326;%%
  %3 citations counted in INSPIRE as of 02 Mar 2014
  
  

%\cite{Hinterbichler:2010es}
\bibitem{Hinterbichler:2010es} 
  K.~Hinterbichler and J.~Khoury,
  %``Symmetron Fields: Screening Long-Range Forces Through Local Symmetry Restoration,''
  Phys.\ Rev.\ Lett.\  {\bf 104}, 231301 (2010).
%  [arXiv:1001.4525 [hep-th]].
  %%CITATION = ARXIV:1001.4525;%%
  %115 citations counted in INSPIRE as of 02 Mar 2014
  
  %\cite{Horndeski:1974wa}
\bibitem{Horndeski:1974wa} 
  G.~W.~Horndeski,
  %``Second-order scalar-tensor field equations in a four-dimensional space,''
  Int.\ J.\ Theor.\ Phys.\  {\bf 10}, 363 (1974).
  %%CITATION = IJTPB,10,363;%%
  %164 citations counted in INSPIRE as of 04 Mar 2014
  
  %\cite{Deffayet:2009mn}
\bibitem{Deffayet:2009mn} 
  C.~Deffayet, S.~Deser and G.~Esposito-Farese,
  %``Generalized Galileons: All scalar models whose curved background extensions maintain second-order field equations and stress-tensors,''
  Phys.\ Rev.\ D {\bf 80}, 064015 (2009).
%  [arXiv:0906.1967 [gr-qc]].
  %%CITATION = ARXIV:0906.1967;%%
  %202 citations counted in INSPIRE as of 04 Mar 2014
  
  %\cite{Nicolis:2008in}
\bibitem{Nicolis:2008in} 
  A.~Nicolis, R.~Rattazzi and E.~Trincherini,
  %``The Galileon as a local modification of gravity,''
  Phys.\ Rev.\ D {\bf 79}, 064036 (2009).
%  [arXiv:0811.2197 [hep-th]].
  %%CITATION = ARXIV:0811.2197;%%
  %476 citations counted in INSPIRE as of 04 Mar 2014
  
  %\cite{Deffayet:2009wt}
\bibitem{Deffayet:2009wt} 
  C.~Deffayet, G.~Esposito-Farese and A.~Vikman,
  %``Covariant Galileon,''
  Phys.\ Rev.\ D {\bf 79}, 084003 (2009).
%  [arXiv:0901.1314 [hep-th]].
  %%CITATION = ARXIV:0901.1314;%%
  %253 citations counted in INSPIRE as of 04 Mar 2014
  
  %\cite{Deffayet:2013lga}
\bibitem{Deffayet:2013lga} 
  C.~Deffayet and D.~A.~Steer,
  %``A formal introduction to Horndeski and Galileon theories and their generalizations,''
  Class.\ Quant.\ Grav.\  {\bf 30}, 214006 (2013).
%  [arXiv:1307.2450 [hep-th]].
  %%CITATION = ARXIV:1307.2450;%%
  %6 citations counted in INSPIRE as of 04 Mar 2014

%\cite{O'Hanlon:1972ya}
\bibitem{O'Hanlon:1972ya} 
  J.~O' Hanlon,
  %``Intermediate-range gravity - a generally covariant model,''
  Phys.\ Rev.\ Lett.\  {\bf 29}, 137 (1972).
  %%CITATION = PRLTA,29,137;%%
  %74 citations counted in INSPIRE as of 03 Mar 2014


%\cite{Teyssandier:1983zz}
\bibitem{Teyssandier:1983zz} 
  P.~Teyssandier and P.~.Tourrenc,
  %``The Cauchy problem for the R+R**2 theories of gravity without torsion,''
  J.\ Math.\ Phys.\  {\bf 24}, 2793 (1983).
  %%CITATION = JMAPA,24,2793;%%
  %156 citations counted in INSPIRE as of 03 Mar 2014
  
  
    %\cite{arXiv:0901.3775}
\bibitem{arXiv:0901.3775} 
  P.~Ho\v rava 
  %``Quantum Gravity at a Lifshitz Point,''
  {\it Phys.\ Rev.\ D\ } {\bf 79} 084008 (2009).
  
      %\cite{Blas:2009qj}
\bibitem{Blas:2009qj}
  D.~Blas, O.~Pujolas and S.~Sibiryakov S,
 {\it Phys.\ Rev.\ Lett.\ } {\bf 104} 181302 (2010).
 
   %\cite{Sotiriou:2010wn}
\bibitem{Sotiriou:2010wn} 
  T.~P.~Sotiriou 
  %``Ho\v rava-Lifshitz gravity: a status report,''
  {\it J.\ Phys.\ Conf.\ Ser.\ } {\bf 283} 012034 (2011).
  %%CITATION = ARXIV:1010.3218;%%
  %71 citations counted in INSPIRE as of 12 Jun 2013
  
  %\cite{Jacobson:2000xp}
\bibitem{Jacobson:2000xp} 
  T.~Jacobson and D.~Mattingly D 
  %``Gravity with a dynamical preferred frame,''
  {\it Phys.\ Rev.\ D} {\bf 64} 024028 (2001).
  
  %\cite{Jacobson:2008aj}
\bibitem{Jacobson:2008aj} 
  T.~Jacobson,
  %``Einstein-aether gravity: A Status report,''
  PoS QG {\bf -PH}, 020 (2007).
%  [arXiv:0801.1547 [gr-qc]].
  %%CITATION = ARXIV:0801.1547;%%
  %102 citations counted in INSPIRE as of 28 Mar 2014
  
  %\cite{Jacobson:2010mx}
\bibitem{Jacobson:2010mx} 
  T.~Jacobson,
  %``Extended Horava gravity and Einstein-aether theory,''
  Phys.\ Rev.\ D {\bf 81}, 101502 (2010)
  [Erratum-ibid.\ D {\bf 82}, 129901 (2010)].
%  [arXiv:1001.4823 [hep-th]].
  %%CITATION = ARXIV:1001.4823;%%
  %58 citations counted in INSPIRE as of 28 Mar 2014
  
  %\cite{Vainshtein:1972sx}
\bibitem{Vainshtein:1972sx} 
  A.~I.~Vainshtein,
  %``To the problem of nonvanishing gravitation mass,''
  Phys.\ Lett.\ B {\bf 39}, 393 (1972).
  %%CITATION = PHLTA,B39,393;%%
  %540 citations counted in INSPIRE as of 28 Mar 2014

%\cite{Babichev:2013usa}
\bibitem{Babichev:2013usa} 
  E.~Babichev and Céd.~Deffayet,
  %``An introduction to the Vainshtein mechanism,''
  Class.\ Quant.\ Grav.\  {\bf 30}, 184001 (2013).
%  [arXiv:1304.7240 [gr-qc]].
  %%CITATION = ARXIV:1304.7240;%%
  %29 citations counted in INSPIRE as of 28 Mar 2014

\bibitem{hawking2} S.~W.~Hawking, {\em Comm. Math. Phys.} {\bf 25}, 167 (1972).

  %\cite{Sotiriou:2011dz}
\bibitem{Sotiriou:2011dz} 
  T.~P.~Sotiriou and V.~Faraoni,
  %``Black holes in scalar-tensor gravity,''
  Phys.\ Rev.\ Lett.\  {\bf 108}, 081103 (2012).
%  [arXiv:1109.6324 [gr-qc]].
  %%CITATION = ARXIV:1109.6324;%%
  %19 citations counted in INSPIRE as of 02 Dec 2013
  
  %\cite{Barausse:2008xv}
\bibitem{Barausse:2008xv} 
  E.~Barausse and T.~P.~Sotiriou,
  %``Perturbed Kerr Black Holes can probe deviations from General Relativity,''
  Phys.\ Rev.\ Lett.\  {\bf 101}, 099001 (2008).
%  [arXiv:0803.3433 [gr-qc]].
  %%CITATION = ARXIV:0803.3433;%%
  %27 citations counted in INSPIRE as of 07 Apr 2014
  
  %\cite{Jacobson:1999vr}
\bibitem{Jacobson:1999vr} 
  T.~Jacobson,
  %``Primordial black hole evolution in tensor scalar cosmology,''
  Phys.\ Rev.\ Lett.\  {\bf 83}, 2699 (1999).
%  [astro-ph/9905303].
  %%CITATION = ASTRO-PH/9905303;%%
  %62 citations counted in INSPIRE as of 03 Dec 2013


%\cite{Horbatsch:2011ye}
\bibitem{Horbatsch:2011ye} 
  M.~W.~Horbatsch and C.~P.~Burgess,
  %``Cosmic Black-Hole Hair Growth and Quasar OJ287,''
  JCAP {\bf 1205}, 010 (2012).
  %[arXiv:1111.4009 [gr-qc]].
  %%CITATION = ARXIV:1111.4009;%%
  %12 citations counted in INSPIRE as of 03 Dec 2013
  
%\cite{Cardoso:2011xi}
\bibitem{Cardoso:2011xi} 
  V.~Cardoso, S.~Chakrabarti, P.~Pani, E.~Berti and L.~Gualtieri,
  %``Floating and sinking: The Imprint of massive scalars around rotating black holes,''
  Phys.\ Rev.\ Lett.\  {\bf 107}, 241101 (2011).
%  [arXiv:1109.6021 [gr-qc]].
  %%CITATION = ARXIV:1109.6021;%%
  %43 citations counted in INSPIRE as of 07 Apr 2014
  
  %\cite{Cardoso:2013fwa}
\bibitem{Cardoso:2013fwa} 
  V.~Cardoso, I.~P.~Carucci, P.~Pani and T.~P.~Sotiriou,
  %``Black holes with surrounding matter in scalar-tensor theories,''
  Phys.\ Rev.\ Lett.\  {\bf 111}, 111101 (2013)
  [arXiv:1308.6587 [gr-qc]].
  %%CITATION = ARXIV:1308.6587;%%
  %2 citations counted in INSPIRE as of 03 Dec 2013
  
  %\cite{Kanti:1995vq}
\bibitem{Kanti:1995vq} 
  P.~Kanti, N.~E.~Mavromatos, J.~Rizos, K.~Tamvakis and E.~Winstanley,
  %``Dilatonic black holes in higher curvature string gravity,''
  Phys.\ Rev.\ D {\bf 54}, 5049 (1996)
  [hep-th/9511071].
  %%CITATION = HEP-TH/9511071;%%
  %98 citations counted in INSPIRE as of 04 Dec 2013
  
    %\cite{Hui:2012qt}
\bibitem{Hui:2012qt} 
  L.~Hui and A.~Nicolis,
  %``A no-hair theorem for the galileon,''
  Phys.\ Rev.\ Lett.\  {\bf 110}, 241104 (2013).
%  [arXiv:1202.1296 [hep-th]].
  %%CITATION = ARXIV:1202.1296;%%
  %8 citations counted in INSPIRE as of 25 Nov 2013
  
  %\cite{Sotiriou:2013qea}
\bibitem{Sotiriou:2013qea} 
  T.~P.~Sotiriou and S.~-Y.~Zhou,
  %``Black hole hair in generalized scalar-tensor gravity,''
  arXiv:1312.3622 [gr-qc].
  %%CITATION = ARXIV:1312.3622;%%
  %4 citations counted in INSPIRE as of 07 Apr 2014
  
  %\cite{Blas:2011ni}
\bibitem{Blas:2011ni} 
  D.~Blas and S.~Sibiryakov,
  %``Horava gravity versus thermodynamics: The Black hole case,''
  Phys.\ Rev.\ D {\bf 84}, 124043 (2011).
%  [arXiv:1110.2195 [hep-th]].
  %%CITATION = ARXIV:1110.2195;%%
  %20 citations counted in INSPIRE as of 07 Apr 2014
  
  %\cite{Barausse:2011pu}
\bibitem{Barausse:2011pu} 
  E.~Barausse, T.~Jacobson and T.~P.~Sotiriou,
  %``Black holes in Einstein-aether and Horava-Lifshitz gravity,''
  Phys.\ Rev.\ D {\bf 83}, 124043 (2011).
%  [arXiv:1104.2889 [gr-qc]].
  %%CITATION = ARXIV:1104.2889;%%
  %29 citations counted in INSPIRE as of 07 Apr 2014
  
  %\cite{Barausse:2012ny}
\bibitem{Barausse:2012ny} 
  E.~Barausse and T.~P.~Sotiriou,
  %``A no-go theorem for slowly rotating black holes in Horava-Lifshitz gravity,''
  Phys.\ Rev.\ Lett.\  {\bf 109}, 181101 (2012)
  [Erratum-ibid.\  {\bf 110}, 039902 (2013)].
%  [arXiv:1207.6370].
  %%CITATION = ARXIV:1207.6370;%%
  %12 citations counted in INSPIRE as of 07 Apr 2014
  
  %\cite{Barausse:2012qh}
\bibitem{Barausse:2012qh} 
  E.~Barausse and T.~P.~Sotiriou,
  %``Slowly rotating black holes in Horava-Lifshitz gravity,''
  Phys.\ Rev.\ D {\bf 87}, 087504 (2013).
%  [arXiv:1212.1334].
  %%CITATION = ARXIV:1212.1334;%%
  %10 citations counted in INSPIRE as of 07 Apr 2014
  
  %\cite{Barausse:2013nwa}
\bibitem{Barausse:2013nwa} 
  E.~Barausse and T.~P.~Sotiriou,
  %``Black holes in Lorentz-violating gravity theories,''
  Class.\ Quant.\ Grav.\  {\bf 30}, 244010 (2013)
  [arXiv:1307.3359 [gr-qc]].
  %%CITATION = ARXIV:1307.3359;%%
  %5 citations counted in INSPIRE as of 07 Apr 2014



\end{thebibliography}
%

\end{document}